# Non-Eulerian behavior of graphitic materials under compression

Ch. Androulidakis[1], E.N. Koukaras[1,2], M. Hadjinicolaou[1,3], C. Galiotis[1,2,4*]

[1]*Institute of Chemical Engineering Sciences, Foundation of Research and Technology-Hellas (FORTH/ICE-HT), Stadiou Street, Platani, Patras, 26504 Greece*
[2]*Nanotechnology and Advanced Materials Laboratory, Department of Chemical Engineering, University of Patras, Patras 26504 Greece*
[3]*School of Science & Technology, Hellenic Open University, Patras, 26222 Greece*
[4]*Department of Chemical Engineering, University of Patras, Patras, 26504 Greece*

[*]Corresponding author: c.galiotis@iceht.forth.gr, galiotis@chemeng.upatras.gr

## ABSTRACT

The mechanical behavior of graphitic materials is greatly affected by the weak interlayer bonding with van der Waals forces for a range of thickness from nano to macroscale. Herein, we present a comprehensive study of the effect of layer thickness on the compression behavior of graphitic materials such as graphene which are fully embedded in polymer matrices. Raman Spectroscopy was employed to identify experimentally the critical strain to failure of the graphitic specimens. The most striking finding is that, contrary to what would be expected from Eulerian mechanics, the critical compressive strain to failure decreases with increase of flake thickness. This is due to the layered structure of the material and in particular the weak cohesive forces that hold the layers together. The plate phenomenology breaks down for the case of multi-layer graphene, which can be approached as discrete single layers weakly bonded to each other. This behavior is modelled here by considering the interlayer bonding (van der Waals forces) as springs in series, and very good agreement was found between theory and experiment. Finally, it will be shown that in the post failure regime multi-layer graphenes exhibit negative stiffness and thus behave as mechanical metamaterials.

# 1. Introduction

Due to its exceptional mechanical properties[1] graphene holds a great promise as a reinforcing agent in polymer nanocomposites. In order to produce nanocomposites that can compete with conventional long fibre composites, few-layer graphene inclusions rather than monolayer graphene are preferable since high volume fractions can thus be attained[2-4]. Previous studies showed that the interlayer bonding of few-layer graphene affects somewhat their overall mechanical performance mainly at high deformations, but still the high stiffness of 1 TPa is retained[5] and fracture strengths of 126 GPa and 101 GPa for bilayer and trilayer[6], respectively, were obtained from nano-identation experiments. Moreover, promising results of multi-layer graphenes as reinforcements of polymeric materials in tension obtained by either mechanical experiments or atomistic simulations[7]. However, the compression behaviour which is very important for structural applications is yet largely unexplored. Another important issue regarding the mechanics of graphene and 2D materials in general is the applicability of continuum mechanics at the nanoscale, and particularly of the plate idealization [8]. The origin of bending rigidity in single layer graphene differs from that of a continuum plate [9, 10] and other methods need to be employed for the estimation of the bending rigidity of 2D materials [11]. Although there are several studies focusing on this subject for single layer graphene [11], there is no analogous work for the case of thicker graphenes which is crucial for the compression behavior of these flakes as well as their mode of failure.

Graphenes of various thicknesses have been examined under uniaxial or biaxial tension[12, 13] using Raman spectroscopy as a tool for monitoring the local strain[14]. The graphene flakes are either supported or fully embedded in polymer matrices, and by bending the polymers the flakes are strained while the mechanical response is monitored mainly by the shift of the position of the 2D and G[12, 15, 16] peaks with the applied strain. This is now a well-established technique for studying graphene under axial deformations for moderate strain levels (~1.5-2%) and various aspects can be studied[17] such as the stress transfer mechanism in graphene/polymer systems[18].

In compression the achievable range of strain using the bending beam technique is sufficient to capture the mechanical behavior of single layer graphene up to failure that occurs by buckling at ~−0.60% and −0.30% strain level for the cases of fully embedded[15] and simply supported[19] 1LG, respectively. As mentioned above there are limited studies for multi-layer graphenes under compression at least experimentally[20, 21], while there have been compression studies of other less ordered graphitic structures such as aerographite[22] and 3D carbon nanotube assemblies[23]. In the present study we examine in detail the compression behavior of simply supported and fully embedded graphene in polymers with thicknesses ranging from bi-layer graphene (2LG) to multi-layer (<10 layers) using Raman spectroscopy and applying continuum theory to acquire an in depth understanding of the failure mechanisms. A comparative reference is also made to the compressive behavior of monolayer graphene that has been examined in our earlier work[15, 16].

**2. Experimental section**

Graphene flakes prepared with mechanical exfoliation of graphite using the scotch tape method[24]. The exfoliated graphitic materials deposited directly on the surface of the substrate PMMA/SU-8 polymer. The SU-8 photoresist was spin coated with speed of ~ 4000 rpm, resulting in a very thin layer of thickness ~180 nm. Appropriate few-layer flakes located with an optical microscope and the number of layers was identified from the 2D Raman peak. In order to create fully embedded flakes another layer of PMMA was spin coated on the top of the flakes with thickness of ~180 nm. A four-point-bending apparatus was used to subject the samples to compressive strain, which was adjusted under the Raman microscope for simultaneously loading the sample and recording spectra. A schematic of the experimental setup is presented in figure S1e. All the experiments performed using a laser line of 785 nm. Strain applied in a stepwise manner and Raman measurements were taken for the 2D and G peaks *in situ* for every strain level. Several points were measured close to the geometric centre of the flakes.

## 3. Results and discussion

*3.1 Experiments*

Here we present experimental results on the compressive behavior of graphene flakes of various thicknesses fully embedded in polymer matrices. Following the setup of previous studies[13, 16, 25], graphene flakes deposited on a PMMA/SU-8 substrate and another layer of thin PMMA was spin coated on the top in order to fully embed the graphene in matrices. Using a four-point-bending jig under a Raman microscope we examined embedded few-layer graphene (with the few-layer we refer to thickness between two and six layers) flakes under compression. The graphene/polymer specimens were subjected to incrementally applied compressive strain while the Raman spectra recorder at every level of loading. It has been explained in detail in previous works[13, 16, 25], that under compression the frequency of the 2D and G phonons shifts to higher wavenumber (phonon hardening) until reaches a peak value, after which downshift of the frequency follows. The maximum strain that corresponds to the peak value of the phonon hardening is the critical strain to failure since the graphene no longer sustain the compressive strain. For every examined specimen several Raman measurements were taken close to the geometric centre of the flakes to avoid edge effects[18].

**Figure 1** shows the position of the 2D peak versus the applied compressive strain. We examined bi-layer, tri-layer and few layer (estimated to be less than ten layers, see SI) specimens in order to assess the compression behavior of multi-layer graphene similar to those employed in graphene nanocomposites[7]. The critical strain to failure for the embedded bilayer is −0.26% (**figure 1a**). This value was consistent for all the examined flakes (see SI). As mentioned earlier, the 2D slope with strain for the bilayer is ~41.4 cm$^{-1}$/%, which is similar to what has been obtained by other workers in the field[2]. The slope for the 2LG is lower than that obtained in the case of 1LG and this indicates that the carbon bonds are not as highly stressed per increment of strain as for 1LG[15]. We note that all the examined flakes have length (width) of over ~15 (10) microns and thus, there are no size effects that might compromise the stress transfer efficiency and the lower shift rate is due to the layered nature of multilayer flakes as has been discussed in detail previously[2-4].

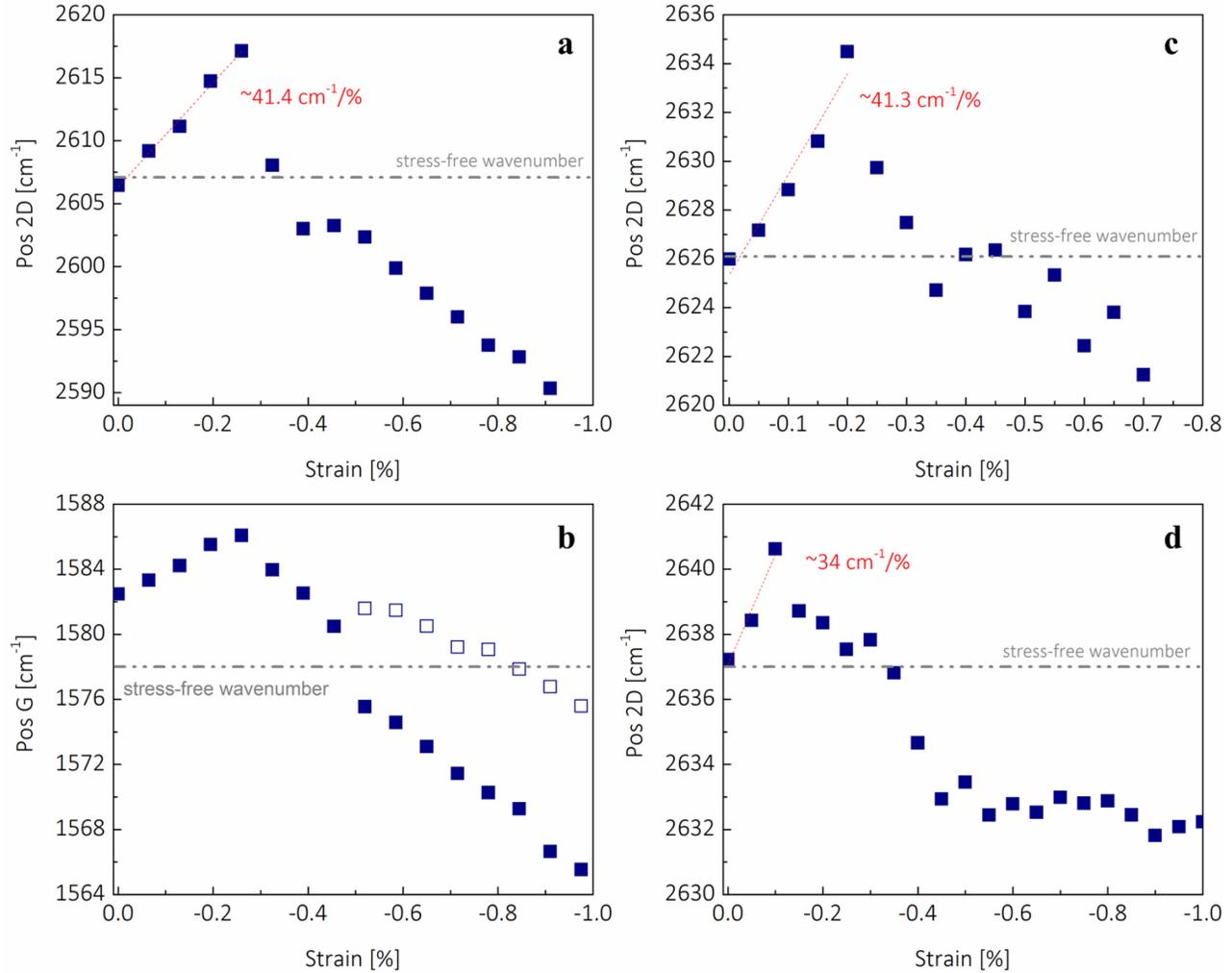

**Fig. 1.** The position of the (a) 2D and (b) G peaks versus the compressive strain for embedded bilayer, and the position of the 2D peak for embedded (c) trilayer and (d) few-layer graphene flakes.

In **figure 1c,d** the results for trilayer and few layer (< 10 layers) graphene specimens are presented. The corresponding critical compressive strain to failure is −0.20% for the trilayer and −0.10% for the few layer, respectively. The slope for both the trilayer and few-layer graphenes is similar to the bilayer and in agreement with results obtained under tension for graphenes with the same thicknesses[3]. We note that the stacking of the few-layers (i.e ABC or ABA) has no effect on the critical strain to failure as revealed by examining flakes of different interlayer configurations (see SI). Moreover, the critical strain to failure is stable for every graphene of same thickness as confirmed by examining more specimens. The results can be found in the SI. The reason for the decrease of the critical compressive strain with the increase of flake thickness must be sought in the way that axial stress (i.e the stress parallel to the direction of the applied tension in the x-axis) is transmitted to multi-layer graphenes. For the embedded flakes the axial stress is transferred from the polymer to the outer layer by shear which is then is transmitted to the inner layers. However, as the inner vdW bonding is much weaker than the polymer/ graphene bond[4] then the interlayer stress transfer is less effective. Thus, a smaller fraction of the total (applied) stress is transferred to the inner layers and an internal shear field is present. As a result, the overall 'structure' is much weaker in compression and fails in shear like a pack of cards under axial compression. This explains why the critical strain to compression failure is significantly smaller than the corresponding value of −0.60% measured previously for monolayer graphene[15]. Further quantitative explanation for this effect is given below.

The post failure behavior is also very interesting. In the case of the trilayer and few layer graphenes the characteristic "slip-stick" behavior is observed in the post failure regime. This is a common occurrence in graphitic materials and has been observed in several studies of nanoscale friction of graphene[26-28]. Another remarkable feature is that after the position of the peaks reaches the zero value (in the post failure regime see **figure 1**), further compression causes downshift of the frequency and clearly passes in the tensile regime. The position of the G peak versus the compressive strain for the bilayer flake is plotted in **figure 1b**. The peak can be fitted by two Lorentzians curves due to the splitting at higher strain level. For compressive strain of ~−0.98% the frequency confirms that the bilayer graphene is in fact under tension (**figure 1b**), and the position of the sub-peaks is ~1575.6 cm$^{-1}$ and ~1565.5 cm$^{-1}$, providing solid evidence that the graphene is under uniaxial tension. In **figure 2** the spectra of the G peak for various levels of compressive strain are presented showing the clear splitting in the tensile regime. In this context the graphene appears to behave as a mechanical metamaterial with negative stiffness. We note that in the work of Tsoukleri et al.[16], the behavior was different in the post failure regime and no such phenomenon was observed, as well as in other samples examined herein and can be found in SI (**figure S1**). The precise mechanism governing this behavior is not perfectly clear, but based on our initial observations we assume that there is a geometric effect for such behaviour to occur. We speculate that a mechanism similar to the incremental negative stiffness behavior observed in carbon nanotubes in the post buckling regime[29] is also the case here for which when the curvature increases tension begins to develop at the curved centre. This is supported by the results of the simply supported flakes, for which the AFM images (see **figure S3**) show buckling failure with localized buckling waves similar to the monolayer[30] while the Raman response has entered the tensile regime. Further examination and simulations are currently under way in order to fully uncover the physical mechanism of this behavior and will be presented in a future study. At any rate, the results here clearly show that graphene exhibits negative stiffness under compression in the post-buckling regime.

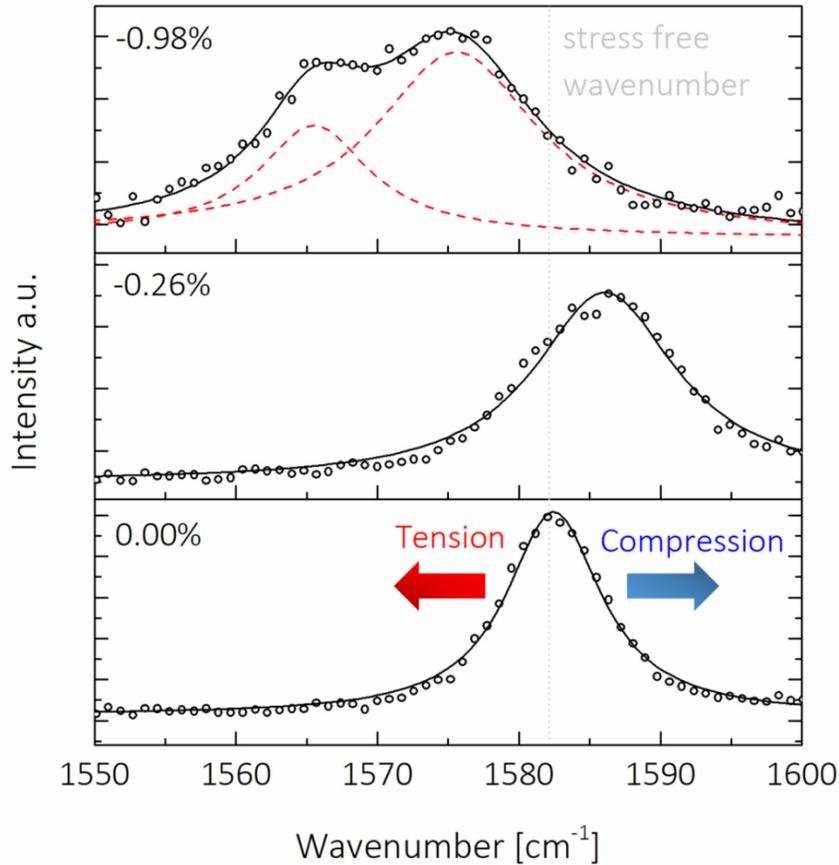

**Fig. 2.** Raman spectra for various levels of compression showing clear splitting at high compression at the tensile regime for a bilayer graphene.

*3.2 Analytical Modelling*

To elucidate the physical mechanism that governs this behavior of fully embedded multi-layer graphitic materials we resort to continuum mechanics. Previously, the compression failure of mono-layer graphene embedded in polymer matrix was modelled using the Winkler's approach [15] for which the interaction between the graphene and the polymer was represented by linear elastic springs and the results were validated by MD simulations [19]. As presented therein, the following set of equations is required to address the problem:

$$\left.\begin{array}{l}\varepsilon_{cr} = \pi^2 \dfrac{D}{C}\dfrac{k}{w^2} + \dfrac{l^2}{\pi^2 C}\left(\dfrac{K_w}{m^2}\right)\\[6pt] k = \left(\dfrac{mw}{l} + \dfrac{l}{mw}\right)^2 \\[6pt] m^2(m+1)^2 = \dfrac{l^4}{w^4} + \dfrac{l^4 K_w}{\pi^4 D}\end{array}\right\} \text{Winkler's model} \qquad (1)$$

Where $\varepsilon_{cr}$ is the critical strain for buckling instability, $D$ and $C$ are the flexural and tension rigidity of the graphene, respectively, $l$ and $w$ are the flake's dimensions (length and width), $k$ is a geometric term, $m$ is the number of the half-waves that the plate buckles, and $K_w$ is the Winkler's modulus (that measures stress per deflection length).

In order to evaluate theoretically the critical strain for Euler buckling instability under compression of embedded graphene flakes that are comprised of more than one layer, the bending and tension rigidities of multi-layer must be known. For the bending rigidity we use low-bound experimentally derived values that are available in the literature [31]. Higher values have also been reported by other workers in the field [10, 32] but as argued below, values of $D$ higher than a certain threshold, lead to larger deviation between standard theoretical approaches and experiment. Furthermore, the same problem holds if we resort to the use of plate mechanics (i.e. by using the well-known formula $D=Eh^3/12(1-v^2)$) for the calculation of the bending rigidity of multi-layer graphene. The tension rigidity is given by $C=Eh$ where $E$ is the stiffness and $h$ the thickness. We assume that the Winkler's modulus does not change with thickness since the foundation support (i.e. the graphene/PMMA bond) is the same. Regarding the values of length and width of the flakes, we use representative values that refer to our experimental specimens bearing in mind that the dimensions do not affect the critical strain to failure [15]. Here, we will examine three cases; (a) the Winkler's model as presented and applied in our previous works [15, 19], using the corresponding bending stiffness values of few-layer flakes that can be found in the literature, and this will be referred to as *Euler buckling*; (b) a spring-in-series model to account for the van der Waals bonds that bind the individual single layers that form the multilayer flakes, and this we will be referred to as *Linear Spring Model*; and finally, (c) a *Modified Winkler* model that retains the linear elastic springs as above and taking the bending rigidity of the multi-layer flake as the product of the bending rigidity of a single layer multiplied with the number of layers. The results of the three cases as above for graphenes with thicknesses of two to six layers are summarized in **Table 1**.

*3.2.1 Euler buckling*

For this case the $K_W$ parameter of reference [15] is employed, since as mentioned above it is unchanged for all cases examined. We observe that this approach yields constant critical strains to buckling with values close to those of the monolayer. This is shown in **figure 3** for fully embedded graphenes under compression and for a wide range of thicknesses. Thus, the predictions for Euler buckling type of failure deviate from the experimental results which clearly indicates the compression behavior of multi-layer and graphitic materials cannot be considered as an elastic buckling phenomenon.

**Table 1.** Experimental results for the critical strain to failure, $\varepsilon_{cr}$, for embedded few-layer graphenes under compression. $K_W$ is the overall Winkler's modulus considering springs in series, $D$ is the bending rigidity for graphenes taken from Ref. [31]. Theoretical estimates of critical strains for the case of Euler buckling (based on the Winkler model presented above), Euler buckling with spring in series, and modified Winkler model, are also presented. In all cases $C=Eh$ as explained in the main text.

| Number of layers | $\varepsilon_{cr}$ (%) Exper. | $K_w$ (springs only between graphene polymer) (GPa/nm) | $K_{w,Tot}$ (spring in series including graphene/graphene interface) (GPa/nm) | $D$[31] (eV) | $D=D_{1L}*n$ (eV) | $\varepsilon_{cr}$ (%) Standard Euler Buckling (Using $K_W$ and $D$) | $\varepsilon_{cr}$ (%) Spring-in-series (Using $K_{W,Tot}$ and $D$) | $\varepsilon_{cr}$ (%) Modified Winkler (Using $K_{W,Tot}$ and $D_{Sl}$) |
|---|---|---|---|---|---|---|---|---|
| 2 | 0.25 | 6 | 1.47 | 3.35 | 2 | 0.52 | 0.27 | 0.21 |
| 3 | 0.20 | 6 | 1.45 | 6.92 | 3 | 0.54 | 0.25 | 0.17 |
| 4 | - | 6 | 1.42 | 12.5 | 4 | 0.54 | 0.25 | 0.14 |
| 5 | - | 6 | 1.40 | 18.1 | 5 | 0.54 | 0.24 | 0.12 |
| 6 | - | 6 | 1.38 | 28.29 | 6 | 0.54 | 0.25 | 0.11 |
| Few-layer | 0.10 | - | - | - | - | - | - | - |

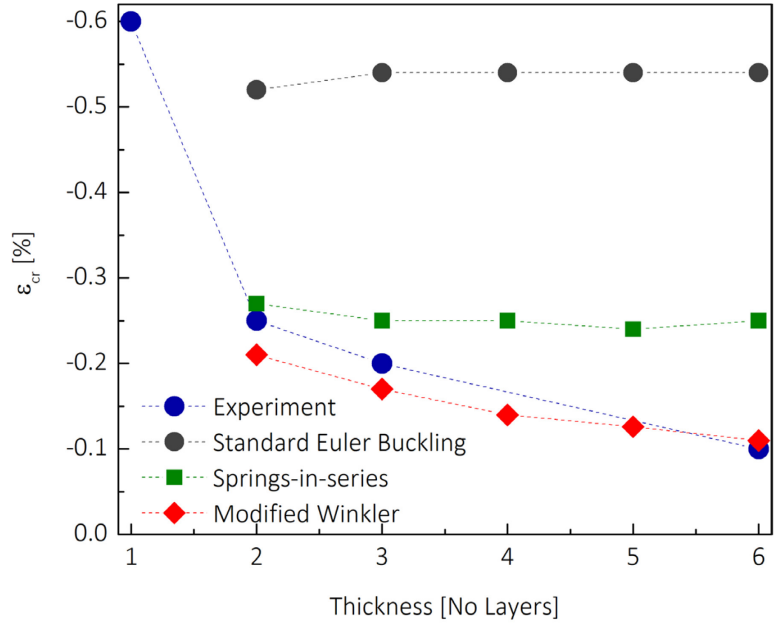

**Fig. 3.** Collective plot of experimental and theoretical results. Blue circles are experimental data, red rhombi are critical strain for failure calculated using the modified Winkler model, green squares are the critical strains obtained theoretically by assuming only springs in series, and triangles represent the strain for Euler buckling form of failure. The dashed lines are guides to the eye.

It seems therefore that the failure of the embedded few layer graphenes in compression is cohesive and originates from the inability of the weak van der Waals bonds to withstand the developed shear stress. It is known that graphite is a solid lubricant with low interlayer shear strength [33, 34]. Loss of stacking has also been observed in embedded few layer graphenes under tension due to relative sliding between individual layers and at relatively small strain[35], which facilitates the cohesive failure of the flakes. Previous studies on the stress transfer mechanism in graphene/polymer systems showed that the shear stress at the interface reached a value of 0.4 MPa at small strains (~0.40%)[18]. The shear strength of graphite has been found to be in a range of 0.25- 2.5 MPa [36, 37] and a value as small as 40 KPa reported for the interlayer shear stress in a bilayer graphene [38]. Thus, the few layer graphenes fail in shear prior to the critical Euler buckling failure. The mono-layer graphene does not suffer from such a structural "disadvantage" and its failure under compression is Euler buckling, as has been discussed in detail previously [15, 19]. More importantly, it possesses the highest resistance to compression among graphitic materials when embedded in polymers despite being the thinnest family member and having the lowest bending rigidity.

*3.2.2 Spring-in-series*

As stated earlier here we introduce a linear -elastic- spring model that binds the individual graphene layers together (**figure 4**). The stress field that is developed during the experiments is also represented in **figure 4**. The mechanical springs that we introduce act as springs in series for the whole system under consideration, i.e. polymer–graphene–polymer. This assumption is also supported by molecular dynamic simulations performed elsewhere [39]. The overall $K_W$ is estimated by considering springs in series (**figure 4**):

$$\frac{1}{K_W} = \frac{1}{K_1} + \frac{n}{K_{gr}} + \frac{1}{K_1} , \qquad (2)$$

where $K_W$ is the Winkler's modulus, $K_1$ is the stiffness of the springs that bind the polymer to the outer layer, $K_{gr}$ is the stiffness of the springs that bond the individual graphene layers to each other with unit of Pa/m, and $n$ is the number of layers of the few layer graphene. The stiffness of the springs that connect the outer graphene layer to the polymer is taken to be $K_1$=3 GPa/nm based on experiments of simply supported mono-layer [19].

The unknown parameter is the stiffness $K_{gr}$ of the springs that connect the individual graphene layers. The value of $K_{gr}$ can be estimated using the interlayer binding energy of graphite or even the cleavage energy. We use the best experimental values currently available of 63.7 meV/atom and 53.97 meV/atom[40] for the cleavage and interlayer binding energy, respectively. To describe the interaction between an individual graphene layer and rest of the layers we consider the latter as a flat substrate surface and follow a well-established approach in the literature[41-43]. In brief, the pairwise Lennard–Jones 12-6 potential,

$$V_{LJ}(r) = \frac{A}{r^{12}} - \frac{B}{r^6},$$

is taken for the interactions between carbon atoms of different layers; atoms are smeared within the layers forming homogenous atom densities; under the additivity assumption a pairwise integration (summation at the continuum limit) of the Lennard–Jones 12–6 potential results in the monolayer–surface interaction energy per unit area, $U_{vdW}$, that has the form of a Lennard–Jones 9–3 potential (following the notation of Aitken et al.[41]):

$$U_{vdW}(z) = \frac{\Gamma_0}{2}\left[\left(\frac{h_0}{z}\right)^9 - 3\left(\frac{h_0}{z}\right)^3\right],$$

where $z$ is the monolayer–surface distance, $h_0$ is the distance when at equilibrium, and $\Gamma_0$ is the interfacial adhesion energy per unit area at equilibrium. The second derivative of the interaction energy leads to the stiffness,

$$k_{vdW}(z) = \frac{9\Gamma_0}{h_0^2}\left[5\left(\frac{h_0}{z}\right)^{11} - 2\left(\frac{h_0}{z}\right)^5\right].$$

Using appropriate values for the interlayer distance and the adhesion energy per unit area, we can estimate the $K_{gr}$ in a range between 87.5 to 198.3 GPa/nm (see SI). Although the range of the values is large, the resulting overall $K_W$ is marginally affected, its value remains almost constant for the range of $K_{gr}$ (figure S4) and consequently the bending rigidity is the crucial parameter for the estimation of the critical strain to failure. Using the $K_W$ estimated considering spring in series (using the value of 87 GPa/nm as representative for the $K_{gr}$) and again using the bending rigidities from reference [31], the obtained critical strain is in closer to the experimental findings but still, the $\varepsilon_{cr}$ seems fixed at ~−0.25% in contrast to the decreasing trend with the increase in thickness and cannot account for the much lower values obtained for the few-layer graphenes.

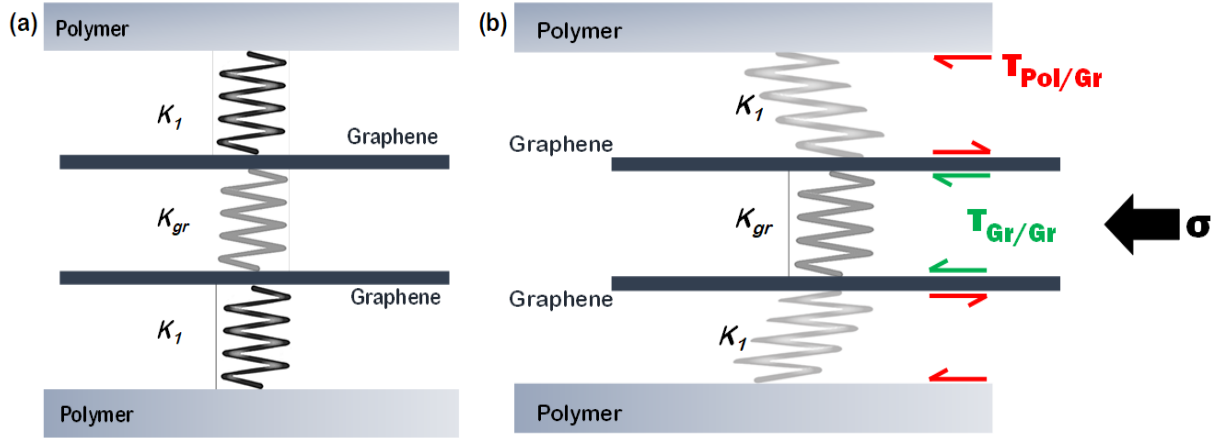

**Fig. 4.** (a) Schematic of the theoretical approach considering that elastic springs of constant $K_1$ bind the graphene with the polymer and corresponding springs of constant $K_{gr}$ binding the internal mono-layers. (b) The developed shear stress fields generated within the layered structure are shown.

*3.2.3 Modified Winkler for multi-layer graphenes*

It has been found that for layered materials, such as $MoS_2$, that are prone to sliding because of the weak interlayer bonding [44], the form of failure under bending differs from that of a continuum plate. In particular, the constituent mono-layers of a given flake bend independently or delaminate, depending on the number of layers [44]. In such cases, the total bending energy of the multi-layer can be estimated by the sum of the bending energy of each individual layer [44]. For the case of multi-layer graphene, this approach has also been supported by earlier theoretical studies [45]. Based on the above, we present herein a Modified Winkler model for which, in addition to the springs-in-series assumption, the bending rigidity, *D*, of multi-layer graphene is given by:

$$D = nD_{1L},\qquad(6)$$

where n is the number of layers of the graphene and $D_{1L}$ is the bending rigidity of a single layer graphene (we use the value of ~ 1 eV for the present calculations[46]). We must note here that a mathematical prerequisite to formulate the theoretical model for the estimation of the critical strain for failure is to define a displacement function that is incorporated to the energy balance of the system. We assign / assume a sinusoidal form for the displacement function[15] which corresponds to the buckling failure of the individuals (single) layers. This however, might not be exactly the case in reality, since interlayer sliding might also occur upon loading. From an energetic point of view, the choice of the displacement function does not affect the critical value in such problems and the present analysis can be applied for the calculation of the critical strain [47]. The results obtained by this approach are plotted in **figure 3** along with the experimental values. As is evident, the modified Winkler model provides a very good agreement to the experimental results. This is not surprising since the weak van der Waals bonds affect greatly the overall mechanical behavior of the few layer graphenes. Another significant finding of the present work is that the shear strength is thickness dependent in this regime, in agreement with previous findings [31].

The spectroscopic data can be converted to values of axial stress (in the x-direction) using the value of 5.5 cm$^{-1}$/GPa for a laser line of 785 nm [48]. Applying the procedure reported previously [48] to the present results, we obtain compressive stress-strain curves for the embedded few layer graphenes. The stress-strain curves are represented in **figure 5** along with results for mono-layer taken from the reference [48] for comparison. A dramatic decrease is observed in the compressive strength with the increase in thickness of the graphenes. The mono-layer has significantly larger compressive strength than thicker graphenes and thus turns up as the most efficient reinforcing filler for composite materials under compressive loadings. Moreover, the change in the sign of stress when the negative stiffness regime is entered are captured.

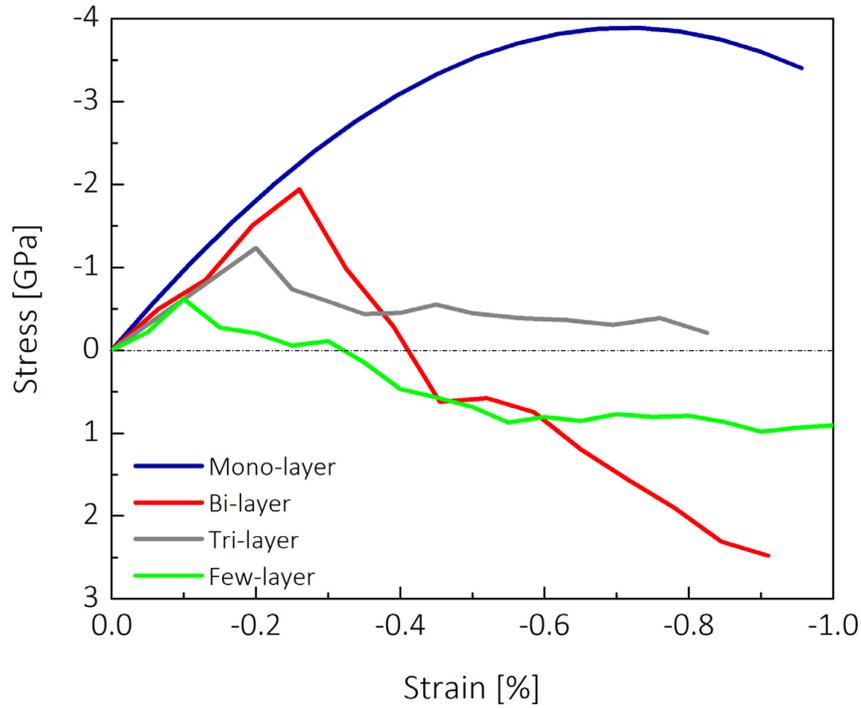

**Fig. 5.** Stress-strain curves for embedded graphenes of various thicknesses under compression derived from the spectroscopic data. The data for the mono-layer were taken from reference [48].

## 4. Conclusions

In summary, we examined in depth the effect of thickness upon the compressive behavior of fully embedded graphenes. The most significant finding is that the critical strain to compression failure decreases with the increase in thickness of graphene as a result of low resistance to shear and consequent cohesive failure. The threshold of cohesive failure occurs at a much lower strain than the calculated critical strain for Euler buckling. The graphene/graphene interactions were treated as springs acting in series, and the overall bending rigidity of a multi-layer graphenes were scaled to the number of layers. This assumption stems from the fact sliding is more favorable failure mechanism than buckling. Finally, the spectroscopic data were converted to stress-strain curves, which showed clearly that the monolayer graphene has a far superior compression behavior than multi-layer graphenes or nano-graphites. These counter-intuitive results are useful for the efficient design of composites that incorporate graphene inclusions as reinforcing agents and provide significant insight for the mechanical behavior of few-layer graphenes.

**Acknowledgements**

The authors acknowledge the financial support of the European Research Council (ERC Advanced Grant 2013) via project no. 321124, "Tailor Graphene". CG also acknowledge the support of of "Graphene Core 2, GA: 696656 – Graphene-based disruptive technologies" which is implemented under the EU-Horizon 2020 Research & Innovation Actions (RIA) and is financially supported by EC-financed parts of the Graphene Flagship.